\documentclass[twocolumn,showpacs,aps,pre,groupedaddress,
amssymb,amsmath,nobalancelastpage]{revtex4}

\usepackage{graphicx}
\usepackage{longtable}

\begin{document}


\title{Impact of boundaries on velocity profiles in bubble rafts}

\author{Yuhong Wang}
\author{Kapilanjan Krishan}
\author{Michael Dennin}
\affiliation{Department of Physics and Astronomy, University of
California at Irvine, Irvine, California 92697-4575}

\date{\today}

\begin{abstract}

Under conditions of sufficiently slow flow, foams, colloids,
granular matter, and various pastes have been observed to exhibit
shear localization, i.e. regions of flow coexisting with regions
of solid-like behavior. The details of such shear localization can
vary depending on the system being studied. A number of the
systems of interest are confined so as to be quasi-two
dimensional, and an important issue in these systems is the role
of the confining boundaries. For foams, three basic systems have
been studied with very different boundary conditions: Hele-Shaw
cells (bubbles confined between two solid plates); bubble rafts (a
single layer of bubbles freely floating on a surface of water);
and confined bubble rafts (bubbles confined between the surface of
water below and a glass plate on top). Often, it is assumed that
the impact of the boundaries is not significant in the
``quasi-static limit'', i.e. when externally imposed rates of
strain are sufficiently smaller than internal kinematic relaxation
times. In this paper, we directly test this assumption for rates
of strain ranging from $10^{-3}$ to $10^{-2}\ {\rm s^{-1}}$. This
corresponds to the quoted rate of strain that had been used in a
number of previous experiments. It is found that the top plate
dramatically alters both the velocity profile and the distribution
of nonlinear rearrangements, even at these slow rates of strain.
When a top is present, the flow is localized to a narrow band near
the wall, and without a top, there is flow throughout the system.

\end{abstract}


\maketitle

\section{Introduction}

When systems are driven sufficiently far from equilibrium, they
often exhibit a series of transitions due to instabilities. This
is particularly common in the flow of fluids, where instabilities
occur at high flow rates. In contrast to this behavior, for
sufficiently {\it slow} driving, complex fluids have been observed
to undergo a transition from a purely flowing state to a
coexistence between a flowing and a solid-like state, i.e. shear
localization
\cite{HBV99,MDKENJ00,LBLG00,DTM01,CRBMGH02,LCD04,HOBRCD05,SCMM03,SBMC03}.
In this context, we focus on complex fluids that are comprised of
dense ``droplets'' (or particles) of one phase or material within
a different continuous phase, such as foams, emulsions, granular
matter, and colloids. We are interested in the case where the
droplets are sufficiently dense that there exists a critical value
of applied stress, the yield stress, below which the material does
not flow at all. In this situation, it has been observed that
under conditions of non-uniform stress the material segregates
into a region that flows (above the yield stress) and a region
that does not flow (below the yield stress) \cite{BOOKS}. However,
because most of these materials are optically opaque, it is only
recently that the spatial dependence of the average velocity of
the ``droplets'' in these materials has been measured
quantitatively. For three dimensional systems, a key development
for such studies has been the development of
magnetic-resonance-imaging \cite{CRBMGH02} techniques that allow
for spatially resolved velocity profiles. Equally useful has been
the use of quasi-two dimensional systems in which all the droplets
can be imaged \cite{DTM01,LCD04}. Coupled with the experimental
advances, there have been a number of simulations that explicitly
look at the possibility of shear localization within the context
of various models of granular matter and foams
\cite{VBBB03,KD03,XOK05a,XOK05b}.

A striking feature of the experimental studies of shear
localization in complex fluids is the division of the velocity
profiles into two basic categories. The first situation
corresponds to cases where the {\it rate of strain is continuous}
across the system \cite{HBV99,MDKENJ00,LBLG00,DTM01}. In this
case, the spatial dependence of the velocity is often exponential.
This appears to be the standard case for granular systems
\cite{HBV99,MDKENJ00,LBLG00} and bubbles confined between two
plates \cite{DTM01}. In contrast, a {\it discontinuity in the rate
of strain} at the transition between the flowing state and the
jammed state is observed in emulsions and colloids
\cite{CRBMGH02}, wet granular systems \cite{HOBRCD05}, worm-like
micelles \cite{SCMM03,SBMC03}, three dimensional foams
\cite{RBC05}, and bubble rafts \cite{LCD04}.

In comparing the systems mentioned above, it is useful to note
that the systems were all sheared between two concentric
cylinders. In this geometry, there is a non-uniform stress across
the system. This might suggest that the localization is due to the
``simple'' picture that part of the system is above the yield
stress and part of the system is below the yield stress.
Surprisingly, there are a number of ways in which the experiments
suggest that this explanation is not sufficient. For example, some
of the systems (especially dry granular systems \cite{LBLG00})
clearly exhibit density variations that impact the flow behavior.
An understanding of these variations is necessary for
understanding the flow localization in these cases. In other
studies, such as with wet granular matter, there are strong
indications that the shear localization is the result of a
viscosity bifurcation \cite{HOBRCD05}.

In contrast to the experiments, simulations have focused on
parallel shear. In this case, a linear velocity transverse to the
shear is expected, and a nonlinear velocity profile is an
indication of some type of shear localization. As with the
experiments, simulations exhibit different behaviors depending on
the details of the model. For example, shear localization is
observed below a critical rate of strain \cite{VBBB03,KD03} and
under different conditions, above a critical rate of strain
\cite{XOK05a}.

For foam the situation is particularly interesting. For
three-dimensional foam, both localized flow \cite{RBC05} and flow
throughout the system \cite{GD99,RCVH03} have been observed. For
quasi-two dimensional experiments, dramatically different types of
flow localization has been observed depending on whether or not
the bubbles were confined between two plates \cite{DTM01} or a
bubble raft was used \cite{LCD04}. These last two experiments
highlight the need for a systematic study of the impact of the
confining plates when studying quasi-two dimensional systems. In
these systems, there is always a lower boundary supporting and
confining the system, and depending on the experiment, there is
often an upper boundary. Typically, the external shear is
generated by motion of the sides, with the upper and lower
boundaries held fixed. Because the focus is understanding the
behavior under conditions of small applied rates of strain, the
systems are often described as being in a quasi-static limit. If
true, the expectation is that the interaction with the confining
boundaries is irrelevant. However, the previous experiments
\cite{DTM01,LCD04} indicate that the boundaries play a critical
role, and suggest that one is not truly in a quasi-static regime,
even though the behavior is rate independent \cite{QUASISTATIC}.

The flow behavior reported on in Refs.~\cite{DTM01,LCD04} used a
Couette geometry, i.e. flow between concentric cylinders. For
bubbles confined between two plates, the shear localization
corresponds to an exponentially decaying velocity as a function of
the distance from the inner cylinder \cite{DTM01}. For the case of
a bubble raft (a single layer of bubbles floating on the surface
of water \cite{BL49,AK79,MGC89}), the velocity as a function of
distance from the inner cylinder exhibited a discontinuity in the
rate of strain \cite{LCD04}. For the case of the confined bubbles,
simulations suggest that nonlinear rearrangements of bubbles
(known as T1 events) provided a focusing of the stress field that
produced the shear localization \cite{KD03}. A T1 event
corresponds to a neighbor switching where two neighboring bubbles
separate, and two bubbles that were not neighbors become neighbors
(see Fig. 1). For the bubble raft, the distribution of T1 events
were studied and no localization was observed \cite{D04}.

As discussed, the most striking difference between the two
experiments is the boundary conditions on the ``top'' and
``bottom'' of the bubbles. The experiments in the confined
geometry have a glass plate in contact with the bubbles both on
the top and the bottom. For the bubble raft, the top surface is
free, as the bubble float on a water surface. There is a third
geometry that has commonly been used to study quasi-two
dimensional foam: a bubble raft with a top plate in contact with
the bubbles. For example, this has been used to study quasi-static
strains \cite{KE99} and the flow around obstacles \cite{DEQRAG05}.
In this paper, we report on experimental studies aimed at
determining the impact of the various boundary conditions. For the
purposes of this comparison, we have focused on relatively
monodisperse systems subjected to parallel shear. Monodisperse
bubbles were used because these systems were the most reproducible
between the two geometries. To allow for minimal variation between
the systems while varying the boundary conditions, we focused on
the two bubble raft systems: with and without a top. For
comparison with past experiments, we consider a range of rate of
strain that was consistent with the rates of strain used in
Ref.~\cite{KD03,LCD04}.

The remainder of the paper is organized as follows. Section II
describes the apparatus and methods for producing the bubble rafts
in detail. Section III describes the method for analyzing the
bubble dynamics, especially the identification of T1 events.
Finally, Sec. IV presents the results and the discussion of the
results.

\begin{figure}
\includegraphics[width=6cm]{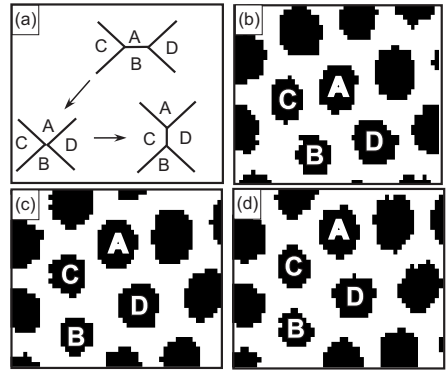}
\caption{\label{t1schem} (a) Schematic representation of a T1
event illustrating two neighboring bubbles (A and B) switching to
next nearest neighbor, and two next nearest neighbors (C and D)
becoming neighbors. (b),(c),(d) A sequence of images that have
been thresholded to bubbles are black taken from a bubble raft
illustrating illustrating a T1 event.}
\end{figure}

\section{Experimental Details}

\begin{figure}
\includegraphics[width=8cm]{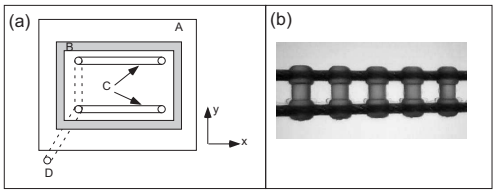}
\caption{\label{schemapp} (a) A schematic of the apparatus as
viewed from the top. The details are described in the text.
Highlighted in the figure are the driving bands (C) that are used
to generate flow. A close up photograph of a driving band is given
in (b). The spacing for the bands is $4\ {\rm mm}$.}
\end{figure}

The experimental setup contains three parts: the trough, the
driving system and the imaging system. A schematic of the trough
is given in Fig.~\ref{schemapp}. The trough consists of a
rectangular Delarin dish (indicated by (A) in the
Fig.~\ref{schemapp}) that is $300\ {\rm mm} \times 400\ {\rm mm}
\times 75\ {\rm mm}$. This serves as the main reservoir for the
aqueous solution. Inside this dish is a Teflon frame (indicated by
(B)) that is used to establish a symmetric boundary and can
support a glass top (not shown in Fig.~\ref{schemapp}). The frame
is held by four poles and controlled by four micrometers outside
of the trough (not shown). The size of the frame is $180\ {\rm mm}
\times 300\ {\rm mm} \times 10\ {\rm mm}$. The frame can move in
three dimensions, with adjustments in the plane used to maintain
symmetric lateral boundaries. The vertical adjustment of the frame
controls the height of the glass top relative to the bubbles.

The bubbles are driven by two counter-rotating belts (indicated by
(C) in Fig.~\ref{schemapp}) using a stepper motor. As indicated in
Fig.~\ref{schemapp}, we define the direction parallel to the belts
to be the x-direction and the direction perpendicular to the belts
as the y-direction. The stepper motor is a Mdrive 23 motor from
Intelligent Motion System, Inc, model number MDMF2222, with
microstepping capability. For driving the foam, the motor is set
to 51200 microstep/rotation. The shafts, gears and belts are from
W.M. Berg Inc. The driving bands are $210\ {\rm mm}$ long and
spaced $57\ {\rm mm}$ apart. All the shafts are mounted at the
bottom of the trough. The shafts are arranged such that the bands
may be driven from outside the trough (through the connection
indicated as (D) in Fig.~\ref{schemapp}). This allows for both
placement of the glass top and imaging the system from above the
glass plate.

The bands that are used to drive the flow act as parallel walls
moving at a constant speed. They are configured to move in
opposite directions, ensuring a location (or region) of zero
velocity in the flowing bubbles. To achieve a no-slip boundary
conditions, belts with a groove spacing on the order of the
average bubble size were used. The top of the belts are set at a
height such that a single row of bubbles fits into the grooves on
the belt.

For imaging the system, a standard CCD camera with a telephoto
lens is used. The lens has a focal length of $6\ {\rm mm}$. The
focus and the aperture are manually adjusted to optimize image
quality by minimizing distortion and balancing the field of view
with magnification of the bubbles. Images from the camera are
directly digitized to the computer using a National Instruments
frame grabber at a maximum frame rate of 30 frames/s. The actual
frame rate was chosen based on the rate of strain to ensure the
ability to track bubble motions. Selecting a frame rate that
corresponded to a total applied strain of 0.001 between images was
found to be adequate to track bubbles without an excessive
overload on the number of images required to analyze sufficiently
long total strains. This requirement combined with the rate of
strain determined the frame rate for any given set of images.

The manufacture of the bubble rafts without a top is discussed in
detail in Ref.~\cite{LTD02}. Essentially, a solution of $80\%$ of
DI water, $15\%$ of glycerin and $5\%$ of a commercially available
bubble solution  (``Miracle bubbles'' from Imperial Toy
Corporation) by volume is used. Compressed nitrogen gas is flowed
through the solution, with the flow rate and needle diameter
controlling the size of the bubbles. By fixing the flow rate, we
were able to generate essentially monodisperse systems. Without a
top, the average diameter of the bubbles was $2.69\ {\rm mm}$,
with a standard deviation of $0.09\ {\rm mm}$ based on fitting the
bubble size distribution to a Gaussian.  The bubble raft is stable
for about two hours without significant popping. For producing the
system with a top, the following procedure was used. A top plate,
made from a 2~mm thick glass, is cleaned with a soap/water
solution. Then, the glass is rinsed thoroughly with the same
solution that constitutes the bubble raft in order to minimize the
influence of any transient wetting or pinning dynamics. The top
glass is placed on the Teflon frame, completely sealing the
system. When making the bubbles, the top plate is moved to one end
of the frame, creating a small opening on the other end. Bubbles
are formed at the closed end, driving the bubbles towards the
opening. When the bubbles fill the entire frame, the glass top is
moved back into position to seal the system. Again, we used a
monodisperse system. With a top, the average diameter of the
bubbles was $2.43\ {\rm mm}$, with a standard deviation of $0.08\
{\rm mm}$ based on fitting the bubble size distribution to a
Gaussian.

\begin{figure}
\includegraphics[width=6cm]{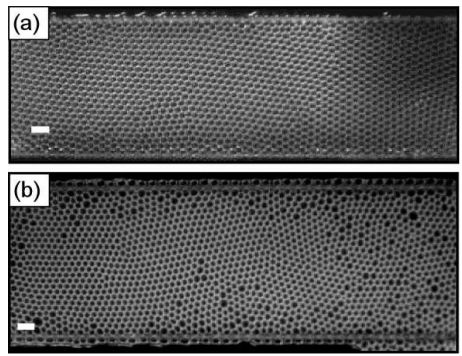}
\caption{\label{images} (a) Image of a typical set of bubbles in
the case with no top on the system. The scale bar represents $7\
{\rm mm}$. (b) Image of a typical set of bubbles for the case of a
top on the system. The scale bar represents $7\ {\rm mm}$.}
\end{figure}

Figure~\ref{images} shows a typical arrangement of the bubbles
with and without the top. Both images include the bands used to
drive the bubbles. One can see that the region outside the bands
is filled with bubbles as well. One other difference between the
two systems is the nature of bubbles well outside the bands. For
the case of a top, because the system is effectively sealed, the
bubbles fill the entire region within the supporting frame. For
the case without a top, the flow outside the bands does show some
unavoidable multilayer formation. This occurs in the corners of
the Teflon frames. The loss of bubbles to these multilayers
results in the formation of voids on the inner perimeter of the
Teflon barriers. The density of bubbles between the bands, in the
region of interest, is not noticeably affected by this. While the
multilayers and void formation may have consequences for the
pressure or stress fields globally, we find our velocity profiles
and T1 rates/densities do not depend on the occurrence or growth
of the multilayers or voids. We have checked for variation along
the x-direction in many of the system properties due to the
influence of the bubbles outside the flowing region. We observe a
small entrance effect that decays rapidly. Therefore, we focus on
the central area of the driven region.

\section{Analysis Methods}

The primary dynamical features of the system we extract are the
velocities of the individual bubbles and local topological
rearrangements. The main topological event of concern for this
paper are the T1 events, where neighbor rearrangements occur (see
Fig.~1).

The raw data from the experiments consists of an image series
capturing the time evolution of the bubbles at different rates of
strain. The analysis of these images may be classified in two
sections (1) A reduction of each image to a set of bubble centers,
edges and vertices and (2) The evolution of these reduced measures
between successive images to extract velocity profiles and T1
events.

The images are initially cropped to a desired region of interest
and Fourier filtered to eliminate noise associated with the CCD
camera and optical non-uniformities. The grayscale images are then
reduced to binary images by thresholding them at an appropriate
value to demarcate the interior regions of bubbles/cells from the
bubble edges. The positions of the centers of each bubble are
computed as the centers of mass of the interior regions of the
cells in such a binary representation. This procedure
reliably identifies over 99\% of the bubbles in each image.

The center positions in consecutive images that show the least
displacement are identified as being associated with the same
bubbles. To reliably make such identification requires the
displacement of the bubbles between successive images be less than
their radii. This was one criteria used in selecting the frame
rate. The velocity of the bubbles is computed using the
displacement of the bubbles between two images and the time taken
for the displacement and averaging over many bubbles and frames.

For the purposes of this paper, the velocity profiles represent an
average over a total applied strain of 5 and a spatial average in
the x-direction. The velocity profile is essentially independent
of the x-position in most of the central region of the trough.
There is a small entrance length at each end in which the velocity
profile varies. Therefore, to be conservative, only the central
1/3 of the trough (in the x-direction) is used for computing
average velocities. To confirm whether or not slip exists at the
driving bands, we computed velocities for the entire width of the
trough (in the y-direction). The y-direction is divided into
evenly spaced bins, and all bubbles in a given y-bin, independent
of their x position, are used to compute the average velocity at
that point.

In our experiments, the the bubbles form densely packed two
dimensional structures. We build a space filling tessellation from
the positions of the cell centers using a Voronoi construction.
The edges and vertices thus extracted are seen to accurately
reproduce the network formed by the edges of individual cells in
the bulk of the system. At the boundaries of the network, the
Voronoi construction is not representative of the cell edges and
therefore in any further analysis we expunge cells that have
vertices at the boundaries.

\begin{figure}
\includegraphics[width=6cm]{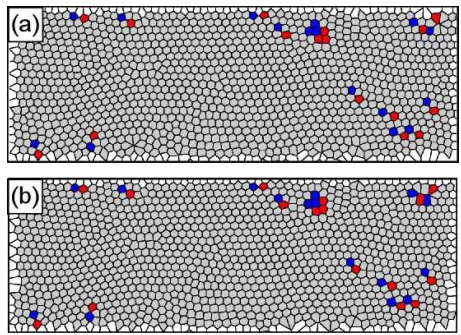}
\caption{\label{voronoi} (color online) Illustration of
identifying T1 events using the Voronoi construction. Image (a)
and image (b) are the Voronoi reconstruction for two sequential
images. Bubbles involved in T1 events are colored.}
\end{figure}

Knowing the vertices shared between cells makes it possible to
identify cells that are neighbors of each other. In order to
identify T1 events occurring in the system, we identify cells for
which the next-nearest neighbors become nearest neighbors. This
scheme identifies two of the cells that participate in a T1-event.
The other two cells correspond to those in which nearest neighbors
become next-nearest neighbors. While this methodology of
identifies pairs of cells participating in T1 events, a number of
such pairs often occur in proximity forming clusters that may be
associated with slip zones. The size of these clusters seen
depends on the framerate of image capture. However, assuming one
had a sufficiently fast camera, all individual T1 events might be
observed. The positions of the T1 events may be computed as the
center of mass of the cells in each cluster. An example of the
Voronoi reconstruction and detection of T1 events is given in
Fig.~\ref{voronoi}. The two images illustrate the system before
and after T1 events occur. Bubbles involved in the T1 events are
shaded for easy identification.

The T1 events correspond to regions where slips between cells
occur resulting in neighbor switching. These events are the
primary mechanisms through which flows in foam systems are known
to occur. The T1 events reflect a variation in the connectivity
between neighboring cells from as a metric independent measure,
while the velocity profiles of the bubbles are based on a
eucledian metric. The relationship between the externally imposed
shear inducing local T1 events and velocity profiles are explored
in the next section.

\section{Results}

\begin{figure}
\includegraphics[width=8.2cm]{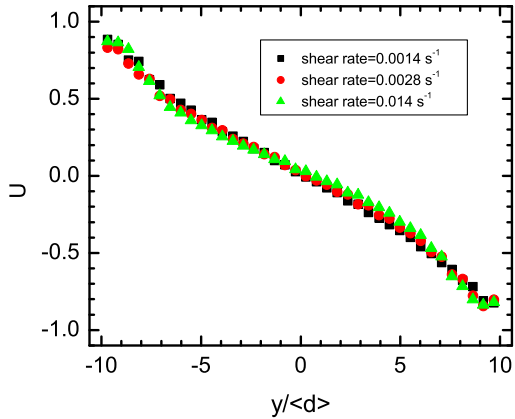}
\caption{\label{velnotop} (color online) The scaled velocity
profile ($U \equiv {\rm (bubble\ velocity)/(belt\ velocity)}$) as
a function of scaled position ($y/<d>$) across the trough for
three different rates of strain for a system without a top, where
$<d>$ is the average bubble diameter. The scale of the position
axis is set so that it extends from the edge of one belt to the
other belt.}
\end{figure}

\begin{figure}
\includegraphics[width=8.2cm]{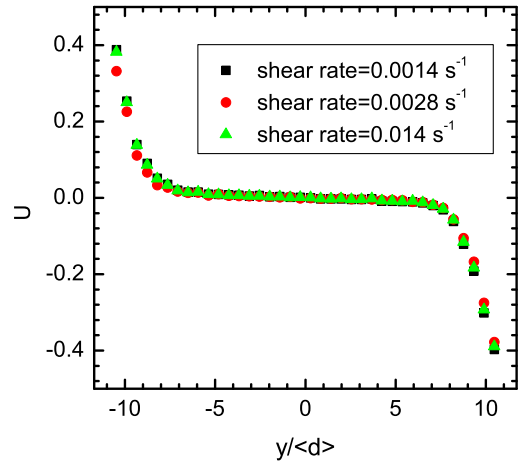}
\caption{\label{veltop} (color online) The scaled velocity profile
($U \equiv {\rm (bubble\ velocity)/(belt\ velocity)}$) as a
function of scaled position ($y/<d>$) across the trough for three
different rates of strain for a system with a top, where $<d>$ is
the average bubble diameter.  The scale of the position axis is
set so that it extends from the edge of one belt to the other
belt.}
\end{figure}

The main result of the paper is a comparison of
Fig.~\ref{velnotop} and Fig.~\ref{veltop}. In both figures, three
different scaled velocity profiles are plotted as a function of
the displacement from the center of the trough in the y-direction
(scaled by the average bubble diameter). (The velocity is scaled
by the driving belt velocity, $U \equiv {\rm (bubble\
velocity)/(belt\ velocity)}$.) A few common features of the
velocity profiles are worth highlighting. If there is no slip at
the boundary, the scaled velocity should be one by definition.
Second, the velocities scale for both boundary conditions and the
three rates of strain reported on here. This indicates that we are
in a rate independent regime \cite{QUASISTATIC}. Finally, because
the bands are moving in opposite directions, the velocity goes
through zero, and it is expected to be zero in the center of the
trough. Both profiles are consistent with this expectation.

\begin{figure}
\includegraphics[width=8.2cm]{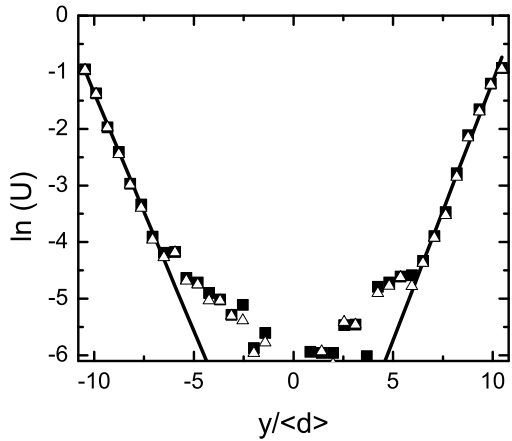}
\caption{\label{exponential} The natural log of the scaled
velocity profile ($\ln (U)$) as a function of scaled position
($y/<d>$) across the trough for two rates of strain in the system
with a top (solid squares are a rate of strain of $0.014\ {\rm
s^{-1}}$ and open triangles are $0.0014\ {\rm s^{-1}}$.) The lines
represent linear fits to the $0.014\ {\rm s^{-1}}$ data,
indicating an exponential decay of the velocity in that regime.}
\end{figure}

The most striking feature is the extreme localization of the flow
when there is a top (Fig.~\ref{veltop}) and the corresponding
essentially linear profile without a top (Fig.~\ref{velnotop}).
This provides strong evidence for the importance of accounting for
the confining boundary conditions, even in a case where one
expects the rate of strain to be sufficiently slow. For comparison
with earlier work, the data with a top is plotted semi-log in
Fig.~\ref{exponential}. One can see that the behavior near each
boundary is consistent with an exponential decay over a few bubble
diameters. The deviations from the exponential behavior in the
center may be in part due to the experimental resolution of our
velocity measurements. Also, it should be noted that the profile
in the case with no top (Fig.~\ref{velnotop}) is not perfectly
linear, as would be expected for a ``simple'' fluid. One candidate
for the deviations from linearity is the monodispersity of the
bubble raft. This is certainly an interesting question, and will
be the subject of future more detailed work. However, for the
purposes of establishing the impact of the boundaries, the
difference between the profiles in Fig.~\ref{velnotop} and
Fig.~\ref{veltop} are more important than the variations from
linear velocity in the case of not having a top.

The other feature of the flow that is apparent is the behavior at
the driving band. In the case of no top, we achieved a no-slip
boundary condition by containing the bubbles in the spaces in the
bands. We tested this by varying the position of the band relative
to the bubbles. If the height of the band was such that the
bubbles sat at the edge of a band but not in one of the gaps, we
observed complete slip at the boundary. In this case, no flow was
observed anywhere in the system. For the case with the top, we
observed some slip at the boundary. However, the ``slip'' was not
complete in the sense that the bands were still able to drive
flow, just with a reduced average speed relative to the speed of
the bands. The introduction of slip in the case of the top is most
likely the result of the drag from the top acting on the bubbles.
An interesting feature of the slip is that the degree of slip was
independent of the rate of strain. This suggests that both the
force between the bubbles and the driving band and the drag of the
plate on the bubbles are independent of rate of strain. One could
test the impact of the plate in the future by selecting driving
bands with varying degrees of interaction between the band and the
bubbles. For a sufficiently strong interaction, one would expect
no slip, despite the drag due to the plate.

\begin{figure}
\includegraphics[width=8.2cm]{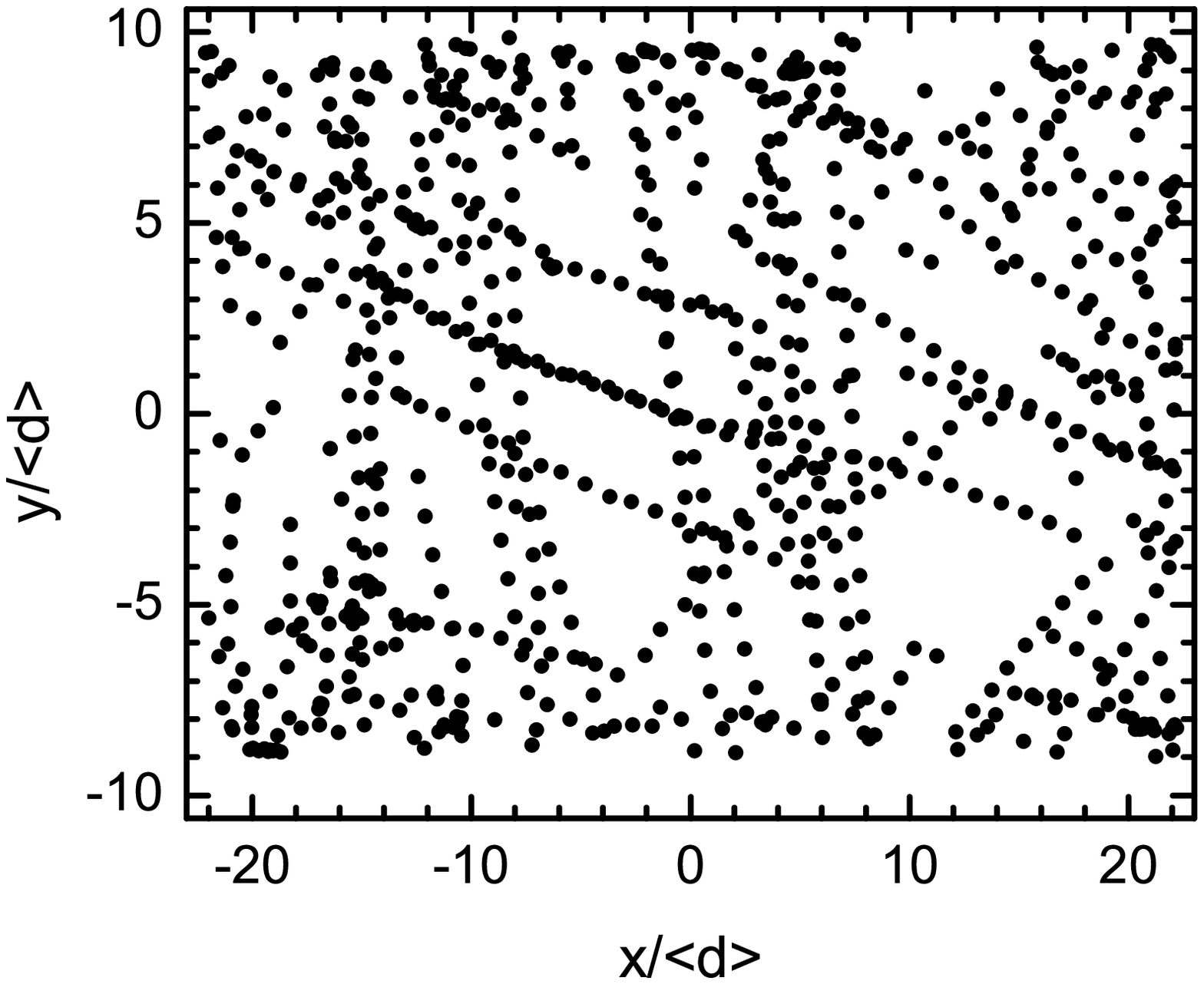}
\caption{\label{t1notop} Positions of T1 events in the central
portion of the system (a fraction of the $x$ direction) without a
confining top. Both the $x$ and $y$ position is scaled by the
average bubble diameter $<d>$.}
\end{figure}

\begin{figure}
\includegraphics[width=8.2cm]{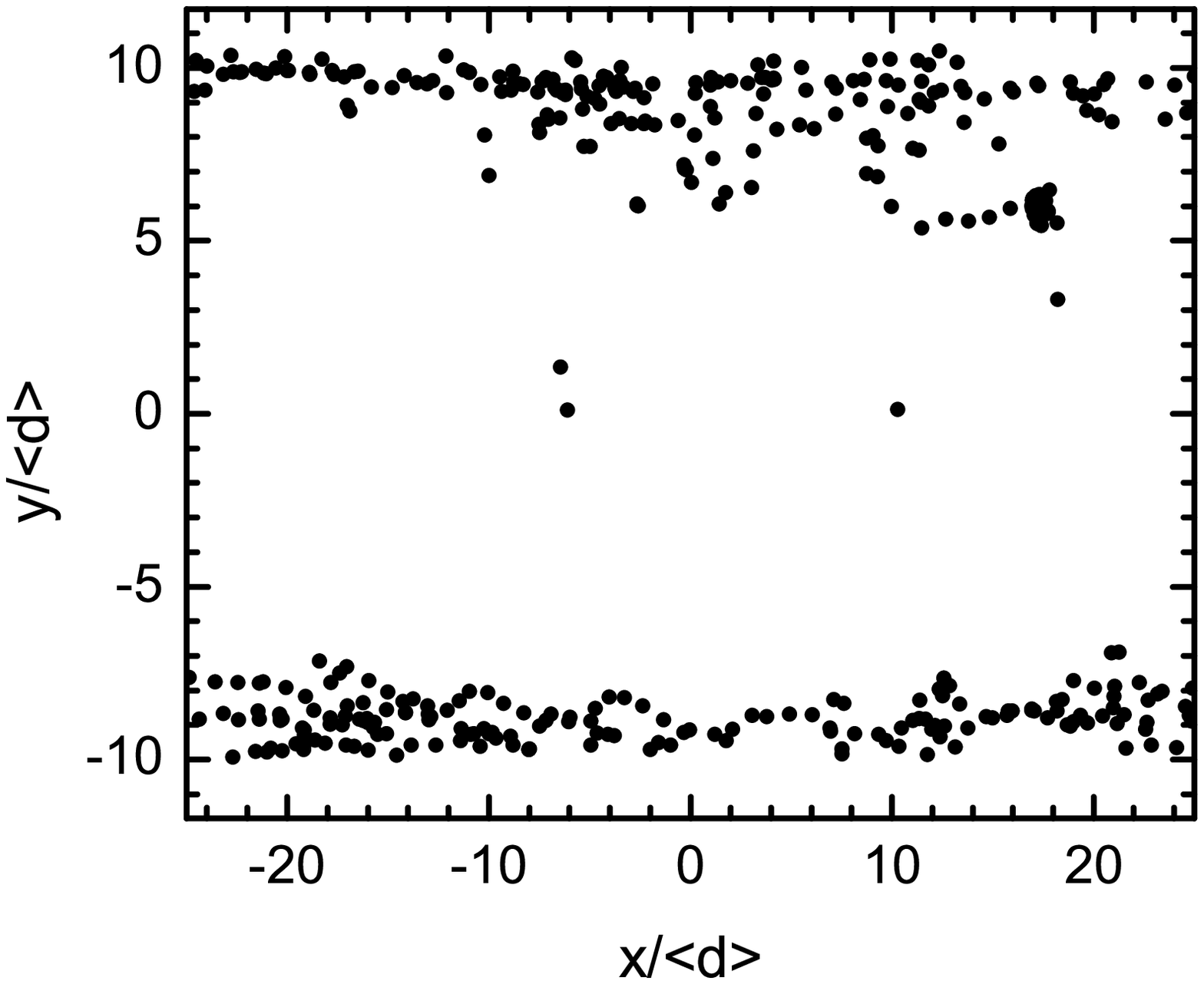}
\caption{\label{t1top}  Positions of T1 events in the central
portion of the system (a fraction of the $x$ direction) without a
confining top. Both the $x$ and $y$ position is scaled by the
average bubble diameter $<d>$.}
\end{figure}

To further explore the impact of the confining top boundary,
Fig.~\ref{t1notop} and \ref{t1top} compare the spatial
distribution of T1 events. For these plots, only a total applied
strain of 0.5 is used. (The smaller interval of strain is used to
avoid overcrowding the plot.) To ensure that the steady state
statistics are being viewed, the last 0.5 of strain out of a total
strain of 5 is selected. Because we are interested in the
differences of the boundary conditions at the slowest possible
rate of strain, only the case for $\dot{\gamma} = 1.4 \times
10^{-3}\ {\rm s^{-1}}$ is shown. Each circle represents the
spatial location of a T1 event.

\begin{figure}
\includegraphics[width=8.2cm]{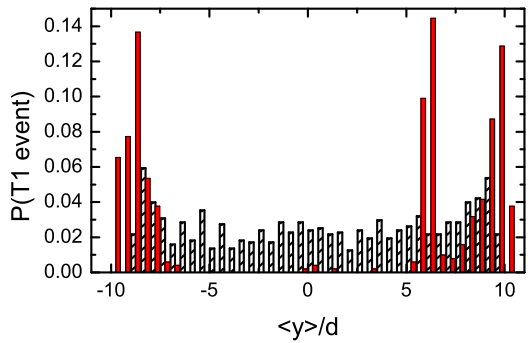}
\caption{\label{t1histo} (Color online) Histogram of the
probability of a T1 event (P(T1 event)) as a function of $y/<d>$
computed using in the range $-22 \leq x/<d>\ \leq 22$ for both the
case with a top (solid red bars) and without a top (hashed open
bars). The histogram shows the absence of T1 events from the
center of the system with a top present, and the relative flatness
of the distribution in the center for flow without a top.}
\end{figure}

The most dramatic feature is the absence of T1 events from the
center region when a top is placed on the system. This is
highlighted in Fig.~\ref{t1histo} where the probability of a T1
event occurring at a particular $y/<d>$ position in the range $-22
\leq x/<d>\ \leq 22$. The histogram illustrates the dramatic
difference between the number of T1 events in the central portion
of the system for the two cases. This indicates the strong
connection between the occurrence of T1 events and the existence
of non-zero velocity. Also of interest, is the slight increase of
T1 events near the boundaries that results in a peak in the
probability for both cases. The peak is close to the boundary, and
more pronounced for the case with the top. The nature of the peaks
reinforces the relative boundary slip for the two cases, as the
probability of T1 events right at the boundary drops to zero for
the case without a top, confirming a lack of slip. In contrast,
with the top, some percentage of T1 events occur even very close
to the boundary, as must happen if slip occurs.

A noteworthy feature of the distribution of T1 events is the
intermittent occurrence of coherent events along lines throughout
the system. These structures appear as lines in Fig.~\ref{t1notop}
and correspond to relaxation through a number of neighbor
rearrangements resulting in large scale slip zones within the
bubble raft. (Various views of a three-dimensional space-time plot
of the occurrence of T1 events is available on EPAPS
\cite{T1CHAINS} that illustrates the temporal correlations between
events.) Preliminary results with a high-speed camera suggests
that the chains of T1 events align with the crystallographic axes
of the bubble lattice, but more systematic work is necessary to
confirm this correlation. Also, it is interesting to speculate on
the correlation between these apparent ``slip planes'' in which
the majority of T1 events occur and the observed systematic
variation in the velocity profile from linear (see
Fig.~\ref{velnotop}). Clearly, future work is needed to both
clarify the impact of monodispersity on the degree of linearity of
the velocity profile and to explore more quantitatively the
connection between T1 events and velocity profiles.

In summary, our results demonstrate that for otherwise identical
systems under conditions of small applied rates of strain, the
existence of a confining solid plate can have a dramatic impact on
the flow behavior of the system by suppressing flow in most of the
system. The suppression of flow is paralleled by a suppression of
T1 events, confirming the expected strong connection between the
two processes. This strongly suggests that even when there is
evidence of rate independence (for example, the scaling of
velocity profiles as seen in this paper), one has to be very
careful in interpreting the measurements from a system with a top
in place.

In trying to understand the role of the top plate, the results
point to three primary sources of damping for the bubble rafts
considered. These correspond to viscous drag between (a) bubbles,
(b) bubbles and the water subphase and (c) bubbles and the glass
top plate (Fig.~\ref{veltop}) or air (Fig.~\ref{velnotop}). The
bubble-bubble interaction can be considered to be the
``intrinsic'' dissipation that provides the effective viscosity
for the system under flow. Previous experiments indicate that the
bubble-bubble interactions are significantly stronger than any
bubble-subphase interaction \cite{LCD04}. The velocity profiles
indicate the relative strength between (a) and (c). With a top,
the exponential decay of velocity at the boundaries
(Fig~\ref{exponential}) suggests that the viscous damping between
the bubbles and the top glass plate dominates relative to the
bubble-bubble interactions, resulting in an exponential profile.
Under the current conditions, without a top, the bubble-bubble
interactions produce a velocity profile that is consistent with a
model that treats the foam as a viscous fluid, even if the
viscosity is non-Newtonian. What remains to be seen is under what,
if any, conditions the velocity profile in the linear case
exhibits significant departures from linear. For example, at
sufficiently slow rates of strain or for larger system sizes, new
behavior may be observed. Finally, it should be noted that recent
work on a system of only two bubbles connects these differences to
whether or not a system is modelled as a dry foam (confining
plates) or a wet foam (bubble raft without a top) \cite{VC05},
which can also be connected with the different mechanisms of
dissipation. This is an interesting connection that requires
further exploration in the large system.

\begin{acknowledgments}

This work was supported by Department of Energy grant
DE-FG02-03ED46071 and PRF 39070-AC9. M. Dennin also thanks the
sponsors and organizers of the Foam Rheology in Two-Dimensions
(FRIT) Workshop held in 2005 at Aberystwyth, Wales.

\end{acknowledgments}


\begin{thebibliography}{28}
\expandafter\ifx\csname
natexlab\endcsname\relax\def\natexlab#1{#1}\fi
\expandafter\ifx\csname bibnamefont\endcsname\relax
  \def\bibnamefont#1{#1}\fi
\expandafter\ifx\csname bibfnamefont\endcsname\relax
  \def\bibfnamefont#1{#1}\fi
\expandafter\ifx\csname citenamefont\endcsname\relax
  \def\citenamefont#1{#1}\fi
\expandafter\ifx\csname url\endcsname\relax
  \def\url#1{\texttt{#1}}\fi
\expandafter\ifx\csname
urlprefix\endcsname\relax\def\urlprefix{URL }\fi
\providecommand{\bibinfo}[2]{#2}
\providecommand{\eprint}[2][]{\url{#2}}

\bibitem[{\citenamefont{Howell et~al.}(1999)\citenamefont{Howell, Behringer,
  and Veje}}]{HBV99}
\bibinfo{author}{\bibfnamefont{D.}~\bibnamefont{Howell}},
  \bibinfo{author}{\bibfnamefont{R.~P.} \bibnamefont{Behringer}},
  \bibnamefont{and} \bibinfo{author}{\bibfnamefont{C.}~\bibnamefont{Veje}},
  \bibinfo{journal}{Phys. Rev. Lett.} \textbf{\bibinfo{volume}{82}},
  \bibinfo{pages}{5241} (\bibinfo{year}{1999}).

\bibitem[{\citenamefont{Mueth et~al.}(2000)\citenamefont{Mueth, Debregeas,
  Karczmar, Eng, Nagel, and Jaeger}}]{MDKENJ00}
\bibinfo{author}{\bibfnamefont{D.~M.} \bibnamefont{Mueth}},
  \bibinfo{author}{\bibfnamefont{G.~F.} \bibnamefont{Debregeas}},
  \bibinfo{author}{\bibfnamefont{G.~S.} \bibnamefont{Karczmar}},
  \bibinfo{author}{\bibfnamefont{P.~J.} \bibnamefont{Eng}},
  \bibinfo{author}{\bibfnamefont{S.~R.} \bibnamefont{Nagel}}, \bibnamefont{and}
  \bibinfo{author}{\bibfnamefont{H.~M.} \bibnamefont{Jaeger}},
  \bibinfo{journal}{Nature} \textbf{\bibinfo{volume}{406}},
  \bibinfo{pages}{385} (\bibinfo{year}{2000}).

\bibitem[{\citenamefont{Losert et~al.}(2000)\citenamefont{Losert, Bocquet,
  Lubensky, and Gollub}}]{LBLG00}
\bibinfo{author}{\bibfnamefont{W.}~\bibnamefont{Losert}},
  \bibinfo{author}{\bibfnamefont{L.}~\bibnamefont{Bocquet}},
  \bibinfo{author}{\bibfnamefont{T.~C.} \bibnamefont{Lubensky}},
  \bibnamefont{and} \bibinfo{author}{\bibfnamefont{J.~P.}
  \bibnamefont{Gollub}}, \bibinfo{journal}{Phys. Rev. Lett.}
  \textbf{\bibinfo{volume}{85}}, \bibinfo{pages}{1428} (\bibinfo{year}{2000}).

\bibitem[{\citenamefont{Debr\'{e}geas et~al.}(2001)\citenamefont{Debr\'{e}geas,
  Tabuteau, and di~Meglio}}]{DTM01}
\bibinfo{author}{\bibfnamefont{G.}~\bibnamefont{Debr\'{e}geas}},
  \bibinfo{author}{\bibfnamefont{H.}~\bibnamefont{Tabuteau}}, \bibnamefont{and}
  \bibinfo{author}{\bibfnamefont{J.~M.} \bibnamefont{di~Meglio}},
  \bibinfo{journal}{Phys. Rev. Lett.} \textbf{\bibinfo{volume}{87}},
  \bibinfo{pages}{178305} (\bibinfo{year}{2001}).

\bibitem[{\citenamefont{Coussot et~al.}(2002)\citenamefont{Coussot, Raynaud,
  Bertrand, Moucheront, Guilbaud, Huynh, Jarny, and Lesueur}}]{CRBMGH02}
\bibinfo{author}{\bibfnamefont{P.}~\bibnamefont{Coussot}},
  \bibinfo{author}{\bibfnamefont{J.~S.} \bibnamefont{Raynaud}},
  \bibinfo{author}{\bibfnamefont{F.}~\bibnamefont{Bertrand}},
  \bibinfo{author}{\bibfnamefont{P.}~\bibnamefont{Moucheront}},
  \bibinfo{author}{\bibfnamefont{J.~P.} \bibnamefont{Guilbaud}},
  \bibinfo{author}{\bibfnamefont{H.~T.} \bibnamefont{Huynh}},
  \bibinfo{author}{\bibfnamefont{S.}~\bibnamefont{Jarny}}, \bibnamefont{and}
  \bibinfo{author}{\bibfnamefont{D.}~\bibnamefont{Lesueur}},
  \bibinfo{journal}{Phys. Rev. Lett.} \textbf{\bibinfo{volume}{88}},
  \bibinfo{pages}{218301} (\bibinfo{year}{2002}).

\bibitem[{\citenamefont{Lauridsen et~al.}(2004)\citenamefont{Lauridsen, Chanan,
  and Dennin}}]{LCD04}
\bibinfo{author}{\bibfnamefont{J.}~\bibnamefont{Lauridsen}},
  \bibinfo{author}{\bibfnamefont{G.}~\bibnamefont{Chanan}}, \bibnamefont{and}
  \bibinfo{author}{\bibfnamefont{M.}~\bibnamefont{Dennin}},
  \bibinfo{journal}{Phys. Rev. Lett.} \textbf{\bibinfo{volume}{93}},
  \bibinfo{pages}{018303} (\bibinfo{year}{2004}).

\bibitem[{\citenamefont{Huang et~al.}(2005)\citenamefont{Huang, Ovarlez,
  Bertrand, Rodts, Coussot, and Bonn}}]{HOBRCD05}
\bibinfo{author}{\bibfnamefont{N.}~\bibnamefont{Huang}},
  \bibinfo{author}{\bibfnamefont{G.}~\bibnamefont{Ovarlez}},
  \bibinfo{author}{\bibfnamefont{F.}~\bibnamefont{Bertrand}},
  \bibinfo{author}{\bibfnamefont{S.}~\bibnamefont{Rodts}},
  \bibinfo{author}{\bibfnamefont{P.}~\bibnamefont{Coussot}}, \bibnamefont{and}
  \bibinfo{author}{\bibfnamefont{D.}~\bibnamefont{Bonn}},
  \bibinfo{journal}{Phys. Rev. Lett.} \textbf{\bibinfo{volume}{94}},
  \bibinfo{pages}{028301} (\bibinfo{year}{2005}).

\bibitem[{\citenamefont{Salmon et~al.}(2003{\natexlab{a}})\citenamefont{Salmon,
  Colin, Manneville, and Molino}}]{SCMM03}
\bibinfo{author}{\bibfnamefont{J.-B.} \bibnamefont{Salmon}},
  \bibinfo{author}{\bibfnamefont{A.}~\bibnamefont{Colin}},
  \bibinfo{author}{\bibfnamefont{S.}~\bibnamefont{Manneville}},
  \bibnamefont{and} \bibinfo{author}{\bibfnamefont{F.}~\bibnamefont{Molino}},
  \bibinfo{journal}{Phys. Rev. Lett.} \textbf{\bibinfo{volume}{90}},
  \bibinfo{pages}{228303} (\bibinfo{year}{2003}{\natexlab{a}}).

\bibitem[{\citenamefont{Salmon et~al.}(2003{\natexlab{b}})\citenamefont{Salmon,
  B\'{e}cu, Manneville, and Colin}}]{SBMC03}
\bibinfo{author}{\bibfnamefont{J.-B.} \bibnamefont{Salmon}},
  \bibinfo{author}{\bibfnamefont{L.}~\bibnamefont{B\'{e}cu}},
  \bibinfo{author}{\bibfnamefont{S.}~\bibnamefont{Manneville}},
  \bibnamefont{and} \bibinfo{author}{\bibfnamefont{A.}~\bibnamefont{Colin}},
  \bibinfo{journal}{Eur. Phys. J. E} \textbf{\bibinfo{volume}{10}},
  \bibinfo{pages}{209} (\bibinfo{year}{2003}{\natexlab{b}}).

\bibitem[{BOO()}]{BOOKS}
\bibinfo{note}{Various books cover both the modelling and experimental
  measurement of yield stress materials, and complex fluids in general. Two
  examples are R. B. Bird, R. C. Armstrong, and O. Hassage, {\it Dynamics of
  Polymer Liquids} (Wiley, New York, 1977) and C. Macosko, {\it Rheology
  Principles, Measurements, and Applications} (VCH Publishers, New York,
  1994).}

\bibitem[{\citenamefont{Varnik et~al.}(2003)\citenamefont{Varnik, Bocquet,
  Barrat, and Berthier}}]{VBBB03}
\bibinfo{author}{\bibfnamefont{F.}~\bibnamefont{Varnik}},
  \bibinfo{author}{\bibfnamefont{L.}~\bibnamefont{Bocquet}},
  \bibinfo{author}{\bibfnamefont{J.-L.} \bibnamefont{Barrat}},
  \bibnamefont{and} \bibinfo{author}{\bibfnamefont{L.}~\bibnamefont{Berthier}},
  \bibinfo{journal}{Phys. Rev. Lett.} \textbf{\bibinfo{volume}{90}},
  \bibinfo{pages}{095702} (\bibinfo{year}{2003}).

\bibitem[{\citenamefont{Kabla and Debr\'{e}geas}(2003)}]{KD03}
\bibinfo{author}{\bibfnamefont{A.}~\bibnamefont{Kabla}} \bibnamefont{and}
  \bibinfo{author}{\bibfnamefont{G.}~\bibnamefont{Debr\'{e}geas}},
  \bibinfo{journal}{Phys. Rev. Lett.} \textbf{\bibinfo{volume}{90}},
  \bibinfo{pages}{258303} (\bibinfo{year}{2003}).

\bibitem[{\citenamefont{Xu et~al.}(2005{\natexlab{a}})\citenamefont{Xu, O'Hern,
  and Kondic}}]{XOK05a}
\bibinfo{author}{\bibfnamefont{N.}~\bibnamefont{Xu}},
  \bibinfo{author}{\bibfnamefont{C.~S.} \bibnamefont{O'Hern}},
  \bibnamefont{and} \bibinfo{author}{\bibfnamefont{L.}~\bibnamefont{Kondic}},
  \bibinfo{journal}{Phys. Rev. Lett.} \textbf{\bibinfo{volume}{94}},
  \bibinfo{pages}{016001} (\bibinfo{year}{2005}{\natexlab{a}}).

\bibitem[{\citenamefont{Xu et~al.}(2005{\natexlab{b}})\citenamefont{Xu, O'Hern,
  and Kondic}}]{XOK05b}
\bibinfo{author}{\bibfnamefont{N.}~\bibnamefont{Xu}},
  \bibinfo{author}{\bibfnamefont{C.~S.} \bibnamefont{O'Hern}},
  \bibnamefont{and} \bibinfo{author}{\bibfnamefont{L.}~\bibnamefont{Kondic}},
  \bibinfo{journal}{cond-mat} p. \bibinfo{pages}{0506507}
  (\bibinfo{year}{2005}{\natexlab{b}}).

\bibitem[{\citenamefont{Rodts et~al.}(2005)\citenamefont{Rodts, , Baudez, and
  Coussot}}]{RBC05}
\bibinfo{author}{\bibfnamefont{S.}~\bibnamefont{Rodts}}, ,
  \bibinfo{author}{\bibfnamefont{J.~C.} \bibnamefont{Baudez}},
  \bibnamefont{and} \bibinfo{author}{\bibfnamefont{P.}~\bibnamefont{Coussot}},
  \bibinfo{journal}{Europhysics Letters} \textbf{\bibinfo{volume}{69}},
  \bibinfo{pages}{636} (\bibinfo{year}{2005}).

\bibitem[{\citenamefont{Gopal and Durian}(1999)}]{GD99}
\bibinfo{author}{\bibfnamefont{A.~D.} \bibnamefont{Gopal}} \bibnamefont{and}
  \bibinfo{author}{\bibfnamefont{D.~J.} \bibnamefont{Durian}},
  \bibinfo{journal}{J. Colloid. Interf. Sci.} \textbf{\bibinfo{volume}{213}},
  \bibinfo{pages}{169} (\bibinfo{year}{1999}).

\bibitem[{\citenamefont{Rouyer et~al.}(2003)\citenamefont{Rouyer, Cohen-Addad,
  Vignes-Adler, and H\"{o}hler}}]{RCVH03}
\bibinfo{author}{\bibfnamefont{F.}~\bibnamefont{Rouyer}},
  \bibinfo{author}{\bibfnamefont{S.}~\bibnamefont{Cohen-Addad}},
  \bibinfo{author}{\bibfnamefont{M.}~\bibnamefont{Vignes-Adler}},
  \bibnamefont{and}
  \bibinfo{author}{\bibfnamefont{R.}~\bibnamefont{H\"{o}hler}},
  \bibinfo{journal}{Phys. Rev. E} \textbf{\bibinfo{volume}{67}},
  \bibinfo{pages}{021405} (\bibinfo{year}{2003}).

\bibitem[{QUA()}]{QUASISTATIC}
\bibinfo{note}{It should be noted that the term ``quasi-static'' is used in
  some work to refer to a regime in which the properties of the system become
  independent of the rate of strain. However, as is demonstrated in
  \cite{TD05}, there is a difference between quasi-static behavior in which the
  system truly relaxes after each step (or the flow is truly slower than any
  relaxation times in the system) and slow, but steady motion that is rate
  independent.}

\bibitem[{\citenamefont{Bragg and Lomer}(1949)}]{BL49}
\bibinfo{author}{\bibfnamefont{L.}~\bibnamefont{Bragg}} \bibnamefont{and}
  \bibinfo{author}{\bibfnamefont{W.~M.} \bibnamefont{Lomer}},
  \bibinfo{journal}{Proc. R. Soc. London, Ser. A}
  \textbf{\bibinfo{volume}{196}}, \bibinfo{pages}{171} (\bibinfo{year}{1949}).

\bibitem[{\citenamefont{Argon and Kuo}(1979)}]{AK79}
\bibinfo{author}{\bibfnamefont{A.~S.} \bibnamefont{Argon}} \bibnamefont{and}
  \bibinfo{author}{\bibfnamefont{H.~Y.} \bibnamefont{Kuo}},
  \bibinfo{journal}{Mat. Sci. and Eng.} \textbf{\bibinfo{volume}{39}},
  \bibinfo{pages}{101} (\bibinfo{year}{1979}).

\bibitem[{\citenamefont{Mazuyer et~al.}(1989)\citenamefont{Mazuyer, Georges,
  and Cambou}}]{MGC89}
\bibinfo{author}{\bibfnamefont{D.}~\bibnamefont{Mazuyer}},
  \bibinfo{author}{\bibfnamefont{J.~M.} \bibnamefont{Georges}},
  \bibnamefont{and} \bibinfo{author}{\bibfnamefont{B.}~\bibnamefont{Cambou}},
  \bibinfo{journal}{J. Phys. France} \textbf{\bibinfo{volume}{49}},
  \bibinfo{pages}{1057} (\bibinfo{year}{1989}).

\bibitem[{\citenamefont{Dennin}(2004)}]{D04}
\bibinfo{author}{\bibfnamefont{M.}~\bibnamefont{Dennin}},
  \bibinfo{journal}{Phys. Rev. E} \textbf{\bibinfo{volume}{70}},
  \bibinfo{pages}{041406} (\bibinfo{year}{2004}).

\bibitem[{\citenamefont{Kader and Earnshaw}(1999)}]{KE99}
\bibinfo{author}{\bibfnamefont{A.~A.} \bibnamefont{Kader}} \bibnamefont{and}
  \bibinfo{author}{\bibfnamefont{J.~C.} \bibnamefont{Earnshaw}},
  \bibinfo{journal}{Phys. Rev. Lett.} \textbf{\bibinfo{volume}{82}},
  \bibinfo{pages}{2610} (\bibinfo{year}{1999}).

\bibitem[{\citenamefont{Dollet et~al.}(2005)\citenamefont{Dollet, Elias,
  Quillet, Raufaste, Aubouy, and Graner}}]{DEQRAG05}
\bibinfo{author}{\bibfnamefont{B.}~\bibnamefont{Dollet}},
  \bibinfo{author}{\bibfnamefont{F.}~\bibnamefont{Elias}},
  \bibinfo{author}{\bibfnamefont{C.}~\bibnamefont{Quillet}},
  \bibinfo{author}{\bibfnamefont{C.}~\bibnamefont{Raufaste}},
  \bibinfo{author}{\bibfnamefont{M.}~\bibnamefont{Aubouy}}, \bibnamefont{and}
  \bibinfo{author}{\bibfnamefont{F.}~\bibnamefont{Graner}},
  \bibinfo{journal}{Phys. Rev. E} \textbf{\bibinfo{volume}{71}},
  \bibinfo{pages}{031403} (\bibinfo{year}{2005}).

\bibitem[{\citenamefont{Lauridsen et~al.}(2002)\citenamefont{Lauridsen,
  Twardos, and Dennin}}]{LTD02}
\bibinfo{author}{\bibfnamefont{J.}~\bibnamefont{Lauridsen}},
  \bibinfo{author}{\bibfnamefont{M.}~\bibnamefont{Twardos}}, \bibnamefont{and}
  \bibinfo{author}{\bibfnamefont{M.}~\bibnamefont{Dennin}},
  \bibinfo{journal}{Phys. Rev. Lett.} \textbf{\bibinfo{volume}{89}},
  \bibinfo{pages}{098303} (\bibinfo{year}{2002}).

\bibitem[{T1C()}]{T1CHAINS}
\bibinfo{note}{See EPAPS Document No. [] for a slide show of different views of
  a space-time plot for a short segment of the data occurence of the T1 events
  given in Fig. 8 of the paper. The vertical axis is time in seconds, and the
  dots are the locations of the T1 events. The space-time plot illustrates the
  temporal correlations of the chains of T1 events. For more information on
  EPAPS, see http://www.aip.org/pubservs/epaps.html.}

\bibitem[{\citenamefont{Vaz and Cox}(2005)}]{VC05}
\bibinfo{author}{\bibfnamefont{M.~F.} \bibnamefont{Vaz}} \bibnamefont{and}
  \bibinfo{author}{\bibfnamefont{S.}~\bibnamefont{Cox}},
  \bibinfo{journal}{Phil. Mag. Lett.} p. \bibinfo{pages}{in press}
  (\bibinfo{year}{2005}).

\bibitem[{\citenamefont{Twardos and Dennin}(2005)}]{TD05}
\bibinfo{author}{\bibfnamefont{M.}~\bibnamefont{Twardos}} \bibnamefont{and}
  \bibinfo{author}{\bibfnamefont{M.}~\bibnamefont{Dennin}},
  \bibinfo{journal}{Physical Review E} \textbf{\bibinfo{volume}{71}},
  \bibinfo{pages}{061401} (\bibinfo{year}{2005}).

\end{thebibliography}

\end{document}